\documentclass[reprint, pre, superscriptaddress]{revtex4-1}

\usepackage{graphicx}

\usepackage{amsmath}

\usepackage{amsfonts}

\usepackage{amssymb}

\usepackage{bm}

\usepackage[caption=false]{subfig}

\usepackage[countmax]{subfloat}

\usepackage{amssymb}
\usepackage{epsf}
\usepackage{pspicture}
\usepackage{flafter}
\usepackage{float}

\begin{document}

\title{Synchronization and phase redistribution in self-replicating populations of coupled oscillators and excitable elements}

\author{Wen Yu}

\author{Kevin B. Wood}
\email{kbwood@umich.edu}
\affiliation{Department of Biophysics, University of Michigan, Ann Arbor, Michigan 48109, USA}
\affiliation{Department of Physics, University of Michigan, Ann Arbor, Michigan 48109, USA}


\begin{abstract}

We study the dynamics of phase synchronization in growing populations of discrete phase oscillatory systems when the division process is coupled to the distribution of oscillator phases.  Using mean field theory, linear stability analysis, and numerical simulations, we demonstrate that coupling between population growth and synchrony can lead to a wide range of dynamical behavior, including extinction of synchronized oscillations, the emergence of asynchronous states with unequal state (phase) distributions, bistability between oscillatory and asynchronous states or between two asynchronous states, a switch between continuous (supercritical) and discontinuous (subcritical) transitions, and modulation of the frequency of bulk oscillations. 

\end{abstract}

\pacs{PACS Nums here}

\maketitle

\section{Introduction}
Synchronization phenomena in collections of coupled oscillators and excitable elements are widely studied in statistical physics~\cite{kuramoto, acebron, strogatzsync, strogatz, kurths}, in part because they represent prototypical nonequilibrium phase transitions exhibiting time-translational symmetry breaking.  In addition to their theoretical value, models of synchronization offer insight into a diverse collection of physical, chemical, and biological phenomena~\cite{kurths,winfree}, ranging from bulk oscillations in chemical reactions to phenotypic or behavioral synchronization in populations of living organisms.  

Synchronization plays a particularly important role in biological systems, where coherent oscillations may serve biological or behavioral function.  Examples abound, including rhythmic flashing of fireflies~\cite{strogatzsync}, coherent behavior of neurons in human neocortex~\cite{grenier} or primate retina~\cite{shlens}, circadian oscillations in cyanobacteria~\cite{Yang:2010aa, david1, david2}, vertebrates~\cite{Yang:2013aa} and mammals~\cite{Reppert:2002aa, Bieler:2014aa}, and synchronized cell division~\cite{segota2014} or protein dynamics~\cite{hasty2010, hasty2014} in populations of single-cell organisms.  In many of these systems, the timescales of the oscillations are well-separated from the timescales of population growth, allowing one to neglect its effects on macroscopic synchrony. In turn, the vast majority of theoretical studies on coupled oscillators have dealt with a fixed population size.  However, this restriction may not always be applicable, as an increasing number of systems have been shown to exhibit oscillations that occur on similar timescales as population growth~\cite{koseka, elowitzPNAS, segota2014, hasty2010, hasty2014, Yang:2010aa, david1, david2}.  This overlap in timescales raises the question of how population growth might affect synchronization, given that the population wide distribution of oscillator phases may be strongly coupled to the growth process.

In this work, we explore the effects of population growth and the corresponding redistribution of phases on the synchronization properties of a simple class of models for both coupled oscillators and excitable elements.  While the paradigmatic Kuramoto model has paved the way to much of our current understanding of synchrony~\cite{acebron, strogatz}, oscillator models with discrete phases have gained increasing attention because of their relative mathematical simplicity~\cite{prager, woodprl,*woodpre1,*woodpre2,*woodpre3, grigolini, cui, rozenblit, assis, assis2, levprl, katja, katja2, katja3, Pinto:2014aa}.  Here, we focus on these discrete phase models because they can be readily modified to integrate oscillator growth and the potential phase dependence of the division and birth processes.  

Using a combination of numerical simulations, mean field theory, and linear stability analysis, we find that the redistribution of phases induced by population growth can disrupt synchronization via either continuous (supercritical) or discontinuous (subcritical) transitions in both discrete phase oscillators and excitable elements.  We observe a range of dynamical behaviors, including bistability between two asynchronous states or between asynchronous and oscillatory states, a switch between supercritical and subcritical transitions as growth is increased, the existence of asynchronous states with unequal phase distributions, or modulation of the bulk oscillation frequency. These results demonstrate that even in minimal models, the coupling between population growth and oscillator phase can profoundly affect synchronization and even lead to new dynamical states that do not exist in the absence of this coupling.

The paper is divided into two sections, with the first devoted to discrete phase oscillators and the second to excitable, discrete phase systems.  In Section~\ref{reviewsec}, we briefly review the discrete phase oscillator model and extend it to capture population growth. In Section~\ref{mf}, we develop a mean field approximation, and in sections~\ref{subseccase1} and~\ref{subseccase2} we use linear stability analysis and numerical simulations to explore the effects of growth when division is independent of state or strongly state-dependent, respectively.  In Section~\ref{reviewexcite} we outline a model for growing populations of excitable elements, in Section~\ref{numsimsE} we describe numerical simulations of the model, and in Section~\ref{mfE} we use mean field theory and linear stability analysis to derive complete phase diagrams for the growing populations.  We conclude with a discussion of the results in Section~\ref{discussion}.

\section{Discrete Phase Oscillators}
\label{oscillators3state}
\subsection{Model for Growing Oscillator Populations}
\label{reviewsec}
We will model an active oscillator as an $m$-state system governed by unidirectional transitions between states, $1 \to 2 \to 3 ...\to m \to 1 \to 2...$~\cite{prager}.  The states represent a type of discretized phase, and formally, the state space of a single oscillator can be described by a phase variable $0 \le \phi \le 2 \pi$ in the limit $m \to \infty$.  For simplicity, we restrict ourselves here to finite $m$ and study a discrete phase oscillator with the minimum number of states ($m=3$) required to generate a Hopf bifurcation and, hence, macroscopic oscillations in a coupled population~\cite{woodprl, *woodpre1,*woodpre2,*woodpre3}.  In addition to their relative simplicity, these discrete phase models are often appropriate descriptions for biological or chemical systems, where the oscillations commonly occur on a discrete state space.

In the absence of coupling, each oscillator transitions irreversibly between states ($1\to2\to3\to1$) with a probability per unit time $g$, which sets the oscillator's intrinsic frequency.  The model is therefore an example of a Markov chain.  Coupling between oscillators is achieved by allowing these transition rates to depend on the fraction of the oscillators in each state; hence, $g$ is replaced by a function $\Gamma_{i}$, which is the probability per unit time for a given oscillator to transition from state $i-1$ to state $i$ (with $i=1,2, 3 \mbox{ (mod 3)}$).   $\Gamma_{i}$ also depends on a real parameter $a \ge 0$, which measures the strength of the coupling between oscillators (hence $\partial \Gamma_i / \partial a \ge 0$).  While a number of nonlinear coupling functions have been used in previous studies~\cite{woodpre3, levprl}, for now we will leave the coupling function unspecified but explicitly note its dependence on the fraction of oscillators $P_{i}$ in state  $i$, $\Gamma_{i}=\Gamma_{i}(P_{i})$.  The primary requirement for this function is that it facilitates phase coherence between oscillators and leads to a Hopf bifurcation at a positive value of $a$.  We discuss these constraints in more detail in what follows.  

In the mean field limit of all-to-all coupling, the fraction of oscillators $P_i$ in state $i$ is governed by the continuous time master equation,
\begin{equation}
\label{mfeqn}
\dot{P}_i = -P_i \Gamma_{i+1} + P_{i-1} \Gamma_{i}.
\end{equation}
The total number of oscillators is conserved in this model ($\sum_i P_i = 1)$, and the fixed point $P_i^*=1/3$ becomes unstable via a Hopf bifurcation as long as 
\begin{equation}
\label{hopf}
\frac{\Gamma_{i}'}{\Gamma_{i}} >3,
\end{equation}
above some critical value of  $a \equiv a_c$~\cite{levprl} .  In Equation~\ref{hopf}, $\Gamma \equiv  \Gamma(x,a) \big|_{x=1/3}$ and $\Gamma_{i}' \equiv \frac{\partial \Gamma(x,a)}{\partial x} \big|_{x=1/3}$.  For larger $a>a_c$, macroscopic oscillations occur.  In what follows, we consider the class of oscillator models where Equation~\ref{hopf} is satisfied for some $a \ge a_c > 0$.

To incorporate population growth, we introduce two minimal mechanisms by which growth and oscillator phase may be coupled.  First, we allow each oscillator in state $i$ to give birth to a new oscillator with probability per unit time $\epsilon_i k$, with $\sum_i \epsilon_i = m = 3$.  The dimensionless weighting factors $\epsilon_i$ couple the rate of division to the state of the oscillator, and $k$ is a growth rate constant.  In this work, we restrict our attention to two limiting cases:  phase independent growth, $\epsilon_i=1$, and strongly phase-dependent growth, $\epsilon_i=3\delta_{ij}$, where $\delta_{ij}$ is the Kronecker delta.  In the former case (Case 1), oscillators in all states are equally likely to divide, so division itself is independent of the state of the oscillator.  In the latter case (Case 2), division can only occur for oscillators in state j.  We choose $j=1$ without loss of generality.  This state dependence of division could be relevant in a number of biological settings, such as synchronization in cell division itself, where the oscillators in certain states--for example, those in certain stages of the cell cycle--divide preferentially.  State-dependence of division could also occur in oscillations of protein dynamics, as transcriptional and translational activity are often linked to the cell cycle.  

Second, we allow the phase of the daughter oscillator to depend probabilistically on the state of the mother.  Specifically, with each division, the new daughter oscillator has probability $\chi$ to be in the same state as the mother and a probability $(1-\chi)/2$ to be in each of the two remaining states.  This assumption could again be relevant in many biological applications, where cell division can lead to a repartitioning of intracellular contents, such as proteins, that may be fundamental to the oscillation.  We stress that our goal here is not to incorporate biological details into a system-specific model, but rather to introduce state-dependent division and birth in a minimal model.  While other rules for coupling single oscillator dynamics to division are possible, we will show below that the simple rules above lead to a rich collection of dynamical behaviors.

\subsection{Mean Field Theory}
\label{mf}
To develop a mean field approximation for this model, we begin by writing evolution equations for $N_i$, the number of oscillators in state $i$, as
\begin{equation}
\label{mfeqn1}
\begin{aligned}
\dot{N}_i&=-N_i \Gamma_{i+1} + N_{i-1} \Gamma_{i} + \\ 
&\epsilon_i k N_i \chi + \frac{k}{2}(\epsilon_{i+1} N_{i+1} + \epsilon_{i-1} N_{i-1})(1-\chi)
\end{aligned}
\end{equation}
with $i=1,2, 3 \mbox{ (mod 3)}$.  The first two terms capture the nonlinear coupling between oscillators that drives synchronization, analogous to the two terms in Equation~\ref{mfeqn}, while the latter two terms account oscillator division, as described in Section~\ref{reviewsec}.  Since the total number of oscillators $\sum_i N_i = N(t)$, we rewrite Equation~\ref{mfeqn1} in terms of $P_i$, the fraction of oscillators in state $i$, using $\dot{P}_i = \dot{N}_i/N - N_i \dot{N} /N^2$, to arrive at
\begin{equation}
\label{generalmf}
\begin{aligned}
\dot{P}_i&=-P_i \Gamma_{i+1} + P_{i-1} \Gamma_{i} + \epsilon_i k P_i \chi +\\ 
& \frac{k}{2}(\epsilon_{i+1} P_{i+1} + \epsilon_{i-1} P_{i-1})(1-\chi)-P_i \frac{\dot{N}}{N},
\end{aligned}
\end{equation}
where $\dot{N}/N = k \sum_j \epsilon_j P_j$ follows directly from Equation~\ref{mfeqn1}.  The model is now fully specified by two differential equations (e.g. for $\dot{P}_1$ and $\dot{P}_2$) and the normalization condition $P_1 + P_2 + P_3 = 1$.  Without loss of generality, we can set $g= 1$, which is equivalent to measuring time in units of $g$ and replacing $k$ with $k/g$; in what follows, we use $k$ for economy of notation.  Equation~\ref{generalmf} provides a general mean field description valid in the limit of all-to-all coupling and can be solved numerically for any choice of parameters.  To make further analytical progress, we now restrict our attention to the limiting cases of phase independent growth, $\epsilon_i=1$, and strongly phase-dependent growth, $\epsilon_i=3\delta_{ij}$.  

\begin{figure}
\includegraphics[width=6.5 cm]{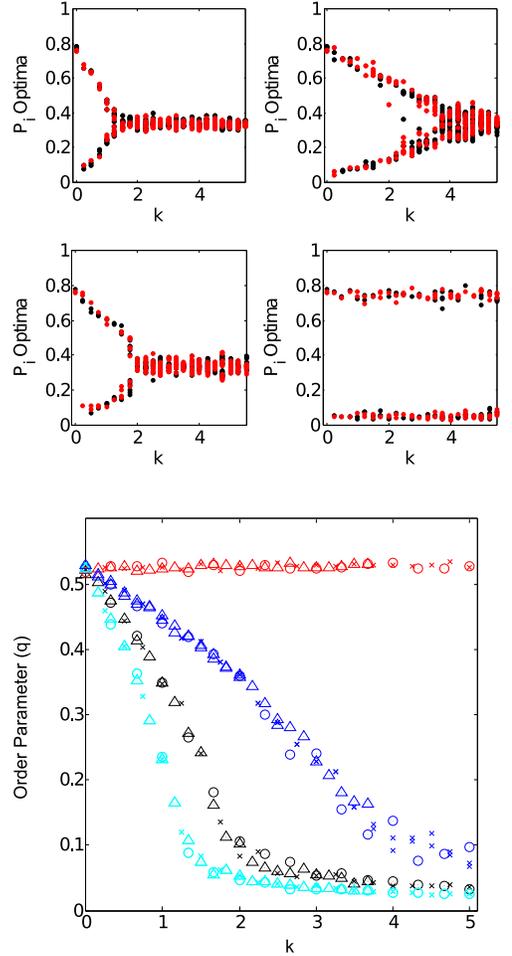}
\caption{Top panels:  values of successive maxima / minima of the time series $P_1(t)$ (black, dark) and $P_2(t)$ (red, light) for each value of the dimensionless growth $k$ for model in Equation~\ref{mfcase1}.  Top left, $\chi=0$; bottom left, $\chi=1/3$; top right, $\chi=2/3$; bottom right, $\chi=1$.  Bottom panel:  order parameter q for $\chi=0$ (light blue), $\chi=1/3$ (black), $\chi=2/3$ (dark blue), and $\chi=1$ (red).  Different shapes represent different intrinsic oscillator frequencies ($g=0.75$, circles; $g=1$, crosses; $g=1.5$, triangles). In all panels, $a=20>a_{c,0}$, $N_0=500$, initial conditions are given by $P_1=2/3$, $P_2=1/3$, and the coupling function $\Gamma_i=1+a P_i^2$.}
\label{case1numericsA}
\end{figure}

\subsection{Case 1:  Division is independent of state}
\label{subseccase1}
When division takes place independently of the state of each oscillator ($\epsilon_i=1$), Equation~\ref{generalmf} reduces to
\begin{equation}
\label{mfcase1}
\begin{aligned}
\dot{P}_i=-P_i \Gamma_{i+1} &+ P_{i-1} \Gamma_{i} - k P_i (1-\chi) +\\ 
& \frac{k}{2}(P_{i+1} + P_{i-1})(1-\chi)
\end{aligned}
\end{equation}
Several things become apparent from inspection.  First, for $\chi=1$ (daughter oscillators are always in the same state as the mother), Equation~\ref{mfcase1} reduces to Equation~\ref{mfeqn}.  In this case, oscillator division leads to an exponentially increasing number of oscillators over time, but all synchronization properties--which depend on the fraction of oscillators in each state, not the total number--will remain unchanged.  Second, it is clear that the asynchronous fixed point $P_i=1/3$ is a solution to Equation~\ref{mfcase1} for all parameter values.  

To analytically explore the effects of division on oscillator synchrony, we linearize around the asynchronous fixed point; the corresponding Jacobian matrix, $J$, evaluated at the fixed point, is
\begin{equation}
\label{jaccase1}
J =  \begin{pmatrix}
-\frac{3}{2} (1-\chi) k - 2 \Gamma + \frac{1}{3}\Gamma'  & -\Gamma - \frac{1}{3} \Gamma'  \\
\Gamma + \frac{1}{3}\Gamma'  & -\frac{3}{2} (1-\chi) k -  \Gamma + \frac{2}{3} \Gamma'   
\end{pmatrix}. 
\end{equation}
where $\Gamma$ and $\Gamma'$ are the coupling function and its derivative, respectively, evaluated at $P_i=1/3$.  The matrix $J$ has complex conjugate eigenvalues $\lambda=\alpha \pm i \omega$, with
\begin{equation}
\begin{aligned}
\alpha &= \frac{1}{2} \left(-3 (1-\chi) k - 3 \Gamma + \Gamma'  \right) \\
\omega &= \frac{1}{2} \left(\sqrt{3} \Gamma + \frac{1}{\sqrt{3}}\Gamma'  \right)  
\end{aligned}
\label{eigvals1}
\end{equation}
which cross the imaginary axis at 
\begin{equation}
k \equiv k_c= \frac{1}{3 (1-\chi)} \left(-3 \Gamma(a_c) + \Gamma'(a_c)  \right),
\label{kcrit}
\end{equation} 
where we have explicitly noted the dependence of $\Gamma$ and $\Gamma'$ on the critical value of the coupling, $a_c$.  Equation~\ref{kcrit} provides a relationship between the critical values of growth $k_c$ and coupling $a_c$, which separate synchronous and asynchronous behavior. For $k=0$, Equation~\ref{kcrit} reduces to Equation~\ref{hopf}.  More generally, for $k>k_c$, the fixed point is stable and no oscillations occur; sufficiently fast population growth therefore disrupts otherwise synchronized populations.  In addition, at the bifurcation point, the frequency of the oscillations is given by $\omega \equiv \omega_0 = \frac{1}{2} \left(\sqrt{3} \Gamma(a_c) + \frac{1}{\sqrt{3}}\Gamma'(a_c)  \right)$ and is therefore expected to be modified by the addition of growth, which changes $a_c$.  

\begin{figure}
\includegraphics[width=7. cm]{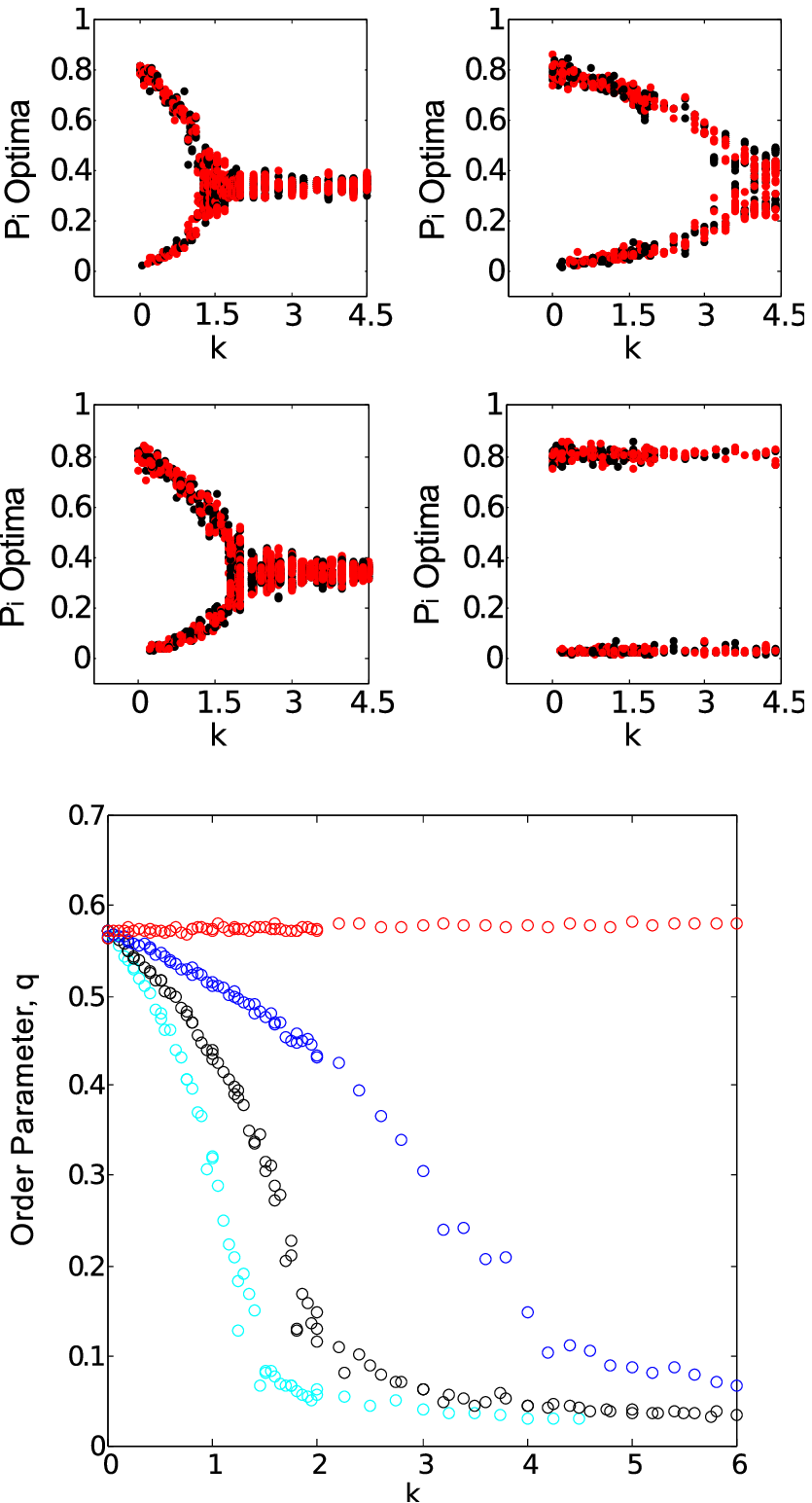}
\caption{Top panels:  values of successive maxima / minima of the time series $P_1(t)$ (black, dark) and $P_2(t)$ (red, light) for each value of the dimensionless growth rate $k$ for model in Equation~\ref{mfcase1}.  Top left, $\chi=0$; bottom left, $\chi=1/3$; top right, $\chi=2/3$; bottom right, $\chi=1$.  Bottom panel:  order parameter q for $\chi=0$ (light blue), $\chi=1/3$ (black), $\chi=2/3$ (dark blue), and $\chi=1$ (red). In all panels, $a=4>a_{c,0}$, $N_0=500$, $g=1$, initial conditions are given by $P_1=2/3$, $P_2=1/3$, and the coupling function $\Gamma_i=e^{a P_i}$.}
\label{case1numericsB}
\end{figure}

\begin{figure}
\includegraphics[width=6.1 cm]{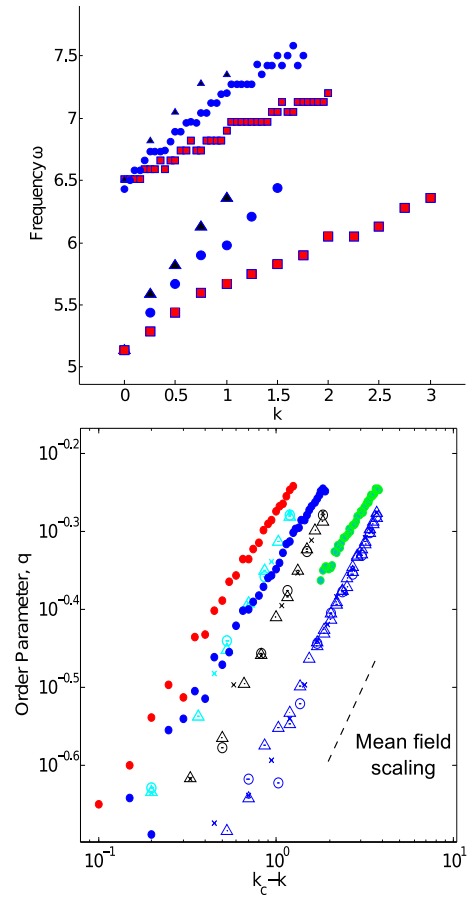}
\caption{Upper panel:  frequency $\omega$ of macroscopic oscillations for model in Equation~\ref{mfcase1}, corresponding to the maximum peak of the power spectrum of $P_1(t)$ for $\chi=0, 1/3, 2/3, 1$ (upper triangles, circles, and squares, respectively).  Curves originating at $\omega \approx 6.5$ correspond to $\Gamma_i=e^{a P_i}$; curves originating at $\omega \approx 5.2$ correspond to $\Gamma_i=1+a P_i^2$.  Lower panel:  order parameter, $q$, as a function of $k_c-k$ near the critical point.  Open shapes correspond to $\Gamma_i=1+a P_i^2$; small closed circles correspond to $\Gamma_i=e^{a P_i}$.  Open shapes: $\chi=0$ (light blue), $\chi=1/3$ (black), $\chi=2/3$ (dark blue) at different intrinsic oscillator frequencies ($g=0.75$, circles; $g=1$, crosses; $g=1.5$, triangles). Closed circles: $g=1$ and $\chi=0$ (blue), $\chi=1/3$ (red), $\chi=2/3$ (green).  In all panels, $N_0=500$ and initial conditions are given by $P_1=2/3$, $P_2=1/3$.  $a=20>a_{c,0}$ for $\Gamma_i=1+a P_i^2$; $a=4>a_{c,0}$ for $\Gamma_i=e^{a P_i}$.  Dashed line: mean field scaling $q \sim (k_c-k)^{1/2}$.  Curves are shifted slightly to allow for visualization of all curves simultaneously. }
\label{case1scaling}
\end{figure}

To determine the nature of the bifurcation (subcritical or supercritical), we calculate the first Lyapunov coefficient, $l_1$.  The sign of $l_1$, which is analogous to the coefficient of the third order term in the normal form for a Hopf bifurcation, is negative for supercritical and positive for subcritical Hopf bifurcations.  For an n-dimensional dynamical system $\dot{x}=f(x,\epsilon)$ with an equilibrium point $x=x^H$ undergoing a Hopf bifurcation at parameter value $\epsilon=\epsilon^H$, $l_1$ can be calculated following the projection procedure in~\cite{Kuznetsov:2004aa} as 
\begin{equation}
\begin{aligned}
l_1 = &\frac{1}{2 \omega_0} Re[\langle p, C(q,q,\bar{q})\rangle - 2 \langle p, B(q,A^{-1} B(q,\bar{q}))\rangle \\
& + \langle p, B(\bar{q},(2 i \omega_0 I-A)^{-1} B(q,q)) \rangle],
\end{aligned}
\end{equation}
where $\langle .,.\rangle$ is the complex scalar product, $I$ is the identity matrix, and $p$ and $q$ are the right and left eigenvectors, respectively, of the Jacobian $A=\frac{\partial f}{\partial x}|_{x=x^H}$ given by
\begin{equation}
\begin{aligned}
A q &= i \omega_0 q, \\
A^T p &= -i \omega_0 p.
\end{aligned}
\end{equation}
The vectors are normalized so that $\langle p,q \rangle=1$, and the functions $B(u,v)$ and $C(u,v,w)$ are multilinear, n-dimensional vector functions given by
\begin{equation}
\begin{aligned}
B(u,v) &= \sum_{j,k=1}^n \frac{\partial^2 f(\psi,\epsilon^H)}{\partial \psi_j \partial \psi_k} \bigg|_{\psi=x^H} u_j v_k \\
C(u,v,w) &= \sum_{j,k,l=1}^n \frac{\partial^3 f(\psi,\epsilon^H)}{\partial \psi_j \partial \psi_k \partial \psi_l} \bigg|_{\psi=x^H} u_j v_k w_l. 
\end{aligned}
\end{equation}
Specifically, for the model in Equation~\ref{mfcase1}, we find
\begin{equation}
\begin{aligned}
q &=\left( \frac{-1-i \sqrt{3}}{2 \sqrt{2}}, \frac{1}{\sqrt{2}} \right), \\
p &=\left(-i \sqrt \frac{2}{3},  -\frac{3 \sqrt{2} + i \sqrt{6}}{6} \right),
\end{aligned}
\end{equation}
independent of the coupling function $\Gamma$.  Following straightforward algebraic manipulations, we arrive at
\begin{equation}
\label{lyap1}
l_1 = \frac{\sqrt 3}{4} \left(\frac{\Gamma''' -6 \Gamma''}{3 \Gamma + \Gamma'} \right).
\end{equation}
where primes indicate derivatives of the coupling function
\begin{equation}
\begin{aligned} 
\Gamma &\equiv  \Gamma(x,a_c) \big|_{x=1/3}, \\
 \Gamma' &\equiv \frac{\partial \Gamma(x,a_c)}{\partial x} \big|_{x=1/3}, \\
  \Gamma'' &\equiv \frac{\partial^2 \Gamma(x,a_c)}{\partial x^2} \big|_{x=1/3},\\
  \Gamma''' &\equiv \frac{\partial^3 \Gamma(x,a_c)}{\partial x^3} \big|_{x=1/3}.
\end{aligned}
\end{equation}
Interestingly, Equation~\ref{lyap1} suggests that the nature of the Hopf bifurcation is determined by the magnitude of the derivatives of the coupling function at the critical point.  Because increasing growth rate, $k$, will change the critical value $a_c$, it is possible for growth to not only change the coupling required for synchronization, but also the nature of the transition itself.  For example, if we make the physically realistic assumptions that $\Gamma_{i} \ge 0$ and $\Gamma'_{i} \ge 0$, the condition $  \Gamma'''=6   \Gamma''$ separates supercritical and subcritical transitions.  To illustrate this point, in the next section we consider two specific examples of the coupling function and show that both supercritical and subcritical bifurcations are possible, depending on its derivatives. 

\begin{figure}
\includegraphics[width=8.75 cm]{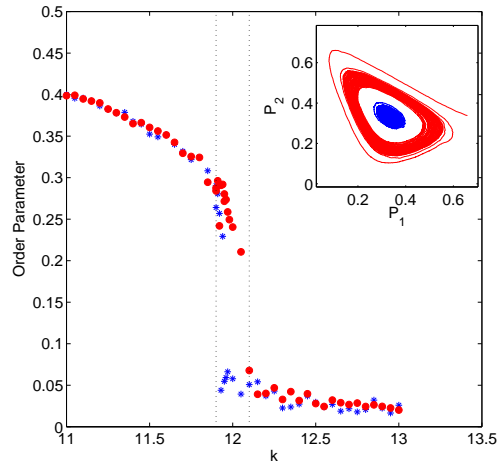}
\caption{Order parameter $q$ as a function of $k$ for Equation~\ref{mfcase1} with $\Gamma_i=e^{a P_i}$ and $a=6.75$. Simulations were run from two sets of initial conditions:  $P_1=2/3, P_2=1/3$ (red) and $P_1=1/3, P_2=1/3$ (blue), leading to different steady state behavior.  Insets: phase portraits for $k=12$ for each set of initial conditions.   $\chi=0$ and $N_0=5000$ for all points.  Dashed lines indicate region of bistability.}
\label{case1bistab}
\end{figure}

\subsubsection{Examples of supercritical and subcritical growth-induced bifurcations}
\label{supersubexamples}
In this section, we study two specific coupling functions to demonstrate the rich dynamics possible in this class of models.  First, we consider a coupling of the form
\begin{equation}
\Gamma(P_i, a) = 1 + a P_i^2,
\end{equation}
which satisfies Equation~\ref{hopf} for $a>a_c \equiv 9$ in the absence of population growth.  For nonzero growth rate $k$, we can rearrange Equation~\ref{kcrit} to show the critical value $a_c$ is increased to
\begin{equation}
a_c = a_{c,0}+ 9 k (1-\chi),
\end{equation} 
where $a_{c,0}$ is the critical coupling in the absence of growth.  In a synchronized population ($a>a_{c,0}$), introducing population growth therefore quashes the macroscopic oscillations when 
\begin{equation}
\label{kcritcase1a}
k>k_c \equiv \frac{a-9}{9(1-\chi)}.
\end{equation}  
At the transition point, the frequency of oscillations is given by
\begin{equation}
\omega_0 = \frac{\sqrt{3}}{6} (a_c+3)  = \sqrt{3} \left(2 + \frac{3}{2} k_c (1-\chi) \right).
\end{equation}
Hence, for a fixed value of $k=k_c$, the frequency at the transition point decreases linearly with $\chi$, eventually reaching the transition frequency $\omega_0=2 \sqrt{3}$ of the non-dividing model.  On the other hand, for a fixed value of $a>a_{c,0}$, the frequency at the transition point approaches the value $\omega=\frac{\sqrt{3}}{6} (a+3)$, independent of $\chi$, as $k$ approaches $k_c$.  

Finally, we can calculate the first Lyapunov coefficient, $l_1$, which depends on $k$, as
\begin{equation}
l_1 = -\frac{9 \sqrt{3} (1+k(1-\chi))}{4+3k(1-\chi)}.
\end{equation} 
Hence, $l_1<0$ for all $k$ and $0 \le \chi \le 1$, indicating that the bifurcation remains supercritical in the presence of population growth.  

As a second example, we consider the coupling
\begin{equation}
\Gamma(P_i, a) = e^{a P_i}.
\end{equation}
In the absence of growth, the model undergoes a Hopf bifurcation at $a=a_{c,0}=3$.  In the presence of growth, the critical value $a_c$ increases and is given by the solution to
\begin{equation}
\label{kcritcase1b}
3 k (1-\chi) = (a-3) e^{a/3}.
\end{equation}
Equivalently, if we start with a synchronized population at a given value of $a>a_{c,0}$, the oscillations will be destroyed when $k>k_c \equiv (a-3) e^{a/3}/(3(1-\chi))$.  The frequency at the transition point is given by
\begin{equation}
\omega_0 = \frac{(a_c+3) e^{a/3}}{2 \sqrt{3}}, 
\end{equation}
which increases when growth is introduced and $a_c>a_{c,0}$. Most interestingly, the first Lyapunov coefficient, $l_1$, is
\begin{equation}
l_1 = -\frac{\sqrt{3} (a_c - 6)a_c^2}{4(3+a_c)},
\end{equation} 
which changes sign at $a_c=6$ (or equivalently, when $\Gamma''' = 6 \Gamma'$).  For small growth rates such that $a_{c,0} < a_c < 6$, the bifurcation between synchronous and asynchronous state is supercritical.  However, for larger growth rates, $a_c > 6$ and the bifurcation becomes a subcritical, discontinuous transition. In this case, population growth--and the corresponding redistribution of oscillator phases--leads to a fundamentally different transition for sufficiently high growth rates.  

To confirm these results numerically, we simulated growing populations of oscillators using the Gillespie algorithm~\cite{gillespie} for a wide range of $k$ (growth rate), $g$ (oscillator natural frequency), and $\chi$ (probability of daughter being in same state as mother).  Unless otherwise noted, simulations were run starting with $N_0=500$ oscillators that are strongly coupled ($a>a_{c,0}$).  For nonzero $k$, the total number of oscillators grows approximately exponentially, so computer memory limits simulations to relatively short time periods.  To circumvent this limitation, we allowed simulations to run until the total number of oscillators reached $N_{max}=10 N_0$; when the number of oscillators exceeded $N_{max}$, we automatically reset the total number of oscillators to $N_0$ while preserving the fractional distribution of oscillator states.  While this numerical procedure effectively underestimates the fluctuations observed in the oscillations, our goal is to approximate a thermodynamic limit $N\to\infty$, and even the modest starting number $N=500$ yields relatively small fluctuations in macroscopic behavior.  

After simulations have reached steady state, we visualize the dynamics by plotting the values of successive maxima / minima of the time series $P_1(t)$ (black, dark) and $P_2(t)$ (red, light) for each value of the dimensionless growth rate $k$ (Figures~\ref{case1numericsA} and ~\ref{case1numericsB}).  We also calculated the synchronization order parameter $q=\left( \langle | Z - \langle Z \rangle_t |^2 \rangle_t \right)^{1/2}$, which was originally proposed in~\cite{grigolini}.  In this definition, $Z(t) = \frac{1}{N} \sum_j e^{i \theta_j(t)}$ is the (not yet averaged) Kuramoto order parameter~\cite{kuramoto, strogatz}, $\theta_j=2 \pi k /3$, $k=(0, 1, 2)$ is the discretized phase of oscillator $j$, and angle brackets represent an average over time.  The standard Kuramoto order parameter is not appropriate for models where rotational symmetry is absent and, consequently, non-oscillating steady states can lead to nonzero values of the order parameter, despite the absence of collective synchrony.  As shown in~\cite{grigolini}, the order parameter $q$ is akin to a generalized standard deviation of $Z(t)$ and removes the bias due to a lack of rotational symmetry.  Nonzero values of $q$ correspond to synchronized, oscillatory states.      

For $\Gamma_i=1+a P_i^2$ at $a=20>a_{c,0}$ (Figure~\ref{case1numericsA}) and $\Gamma_i=e^{a P_i}$ at $a=4>a_{c,0}$ (Figure ~\ref{case1numericsB}), oscillations smoothly decrease in amplitude with increasing $k$, leading eventually to a completely asynchronous state.  The value of the critical growth $k$ is consistent with the linear stability analysis, Equation~\ref{kcritcase1a} and Equation~\ref{kcritcase1b}; it increases with $\chi$ until, at $\chi=1$, the oscillations are undisturbed by even large growth rates.  

Our numerical simulations indicate that population growth affects not only global synchronization, but also the frequency of the macroscopic oscillations in the synchronized state.  To explore this frequency dependence systematically, we calculated the power spectrum of the time series $P_1(t)$ for each simulation in steady state.  Figure~\ref{case1scaling} (top panel) shows the frequency of the dominant peak in the power spectrum, which characterizes the frequency of the macroscopic oscillations.  In both examples, increasing $k$ leads to a monotonically increasing oscillation frequency until, at sufficiently high values of $k$, the macroscopic oscillations are no longer present.  

Because these models are globally coupled and exhibit supercritical bifurcations, one expects that the order parameter should scale as $q \sim (k_c-k)^\beta$ near the critical point, with $\beta=1/2$ the standard mean field scaling exponent.  Indeed, we find that in our numerical simulations, the order parameter approximately follows mean field scaling near the critical point, independent of the value of $\chi$ or the specific coupling function chosen (Figure~\ref{case1scaling}, bottom panel; note that we have slightly shifted the curves relative to one another so that each is visible).  One  would expect similar behavior for any choice of $\Gamma_i$ where a supercritical transition occurs as $k$ eclipses a critical value $k_c$.

By contrast, the transition becomes discontinuous when $\Gamma_i=e^{a P_i}$ and $a$ is increased beyond $a=6$.  For example, at $a=6.75$, growth induces a subcritical Hopf bifurcation (Figure ~\ref{case1bistab}), as indicated by the discontinuous drop in the order parameter and the corresponding bistability.  Interestingly, in the bistable region, which we find to exist for $11.9 \le k \le 12.1$, synchronous oscillations co-exist with an asynchronous fixed point.  For populations initiated with approximately uniform phase distributions, the system settles into a stable asynchronous state; for highly non-uniform distributions, the population undergoes stable oscillations indicative of synchrony (upper right inset).  Similar behavior would be expected in the class of models where $l_1$ can switch signs as growth increases (see Equation~\ref{lyap1}), indicating that in this class of models, coupling between population growth and phase can modify the nature of the Hopf bifurcation.  

\begin{figure}
\includegraphics[width=7.25 cm]{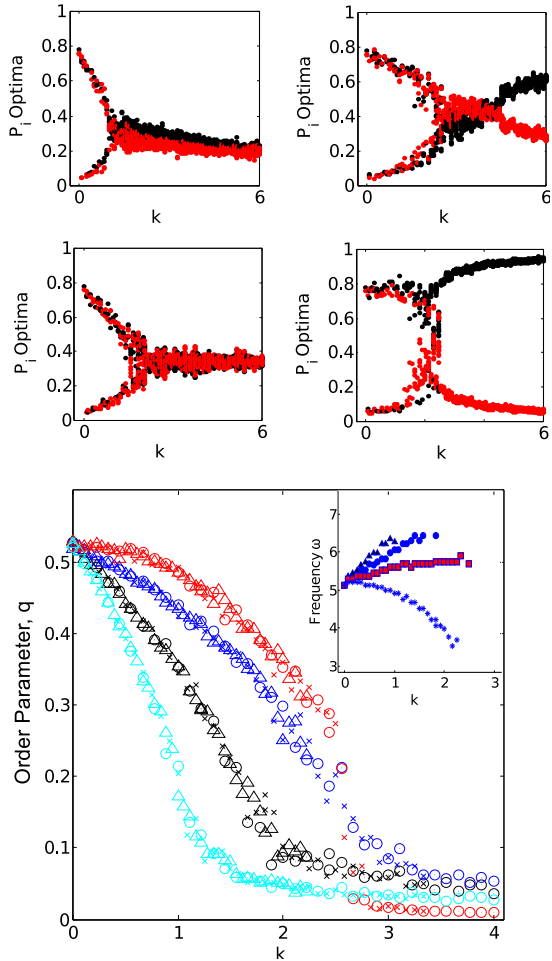}
\caption{Top panels:  values of successive maxima / minima of the time series $P_1(t)$ (black, dark) and $P_2(t)$ (red, light) for each value of the dimensionless ratio $k$ for the model Equation~\ref{mfcase2}.  Top left, $\chi=0$; bottom left, $\chi=1/3$; top right, $\chi=2/3$; bottom right, $\chi=1$.  Bottom panel:  order parameter q for $\chi=0$ (light blue), $\chi=1/3$ (black), $\chi=2/3$ (dark blue), and $\chi=1$ (red).  Different shapes represent different intrinsic oscillator frequencies ($g=0.75$, circles; $g=1$, crosses; $g=1.5$, triangles). Inset:  frequency $\omega$ of macroscopic oscillations, corresponding to the maximum peak of the power spectrum of $P_1(t)$ for $\chi=0, 1/3, 2/3, 1$ (upper triangles, circles, squares and stars, respectively).  In all panels, $a=20>a_{c,0}$, $N_0=500$, initial conditions are given by $P_1=2/3$, $P_2=1/3$, and the coupling function $\Gamma_i=1+a P_i^2$.}
\label{case2numericsA}
\end{figure}

\subsection{Case 2:  Division occurs only in one state}
\label{subseccase2}
When division occurs only in one state ($\epsilon_i=3\delta_{i1}$), equation~\ref{generalmf} reduces to
\begin{equation}
\label{mfcase2}
\begin{aligned}
\dot{P}_i=-P_i \Gamma_{i+1} &+ P_{i-1} \Gamma_{i} +  3 k P_i  ( \delta_{i,1} \chi- P_1) +\\ 
& \frac{3 k}{2}(\delta_{i,3} P_{i+1} + \delta_{i,2} P_{i-1})(1-\chi)
\end{aligned}
\end{equation}
where $\delta_{ij}$ is the Kronecker delta.  Because the choice of ${\epsilon_i}$ has broken the rotational symmetry of the model, the solution $P_i = 1/3$ is not, in general, a steady state solution to Equation~\ref{mfcase2}.  In what follows, we first consider the case $\chi=1/3$, which is amenable to analytical treatment, and then go on to numerically explore the general case $0\le\chi\le1$.

When $\chi=1/3$, Equation~\ref{mfcase2} reduces to
\begin{equation}
\label{mfcase2sc}
\begin{aligned}
\dot{P}_i=-P_i \Gamma_{i+1} &+ P_{i-1} \Gamma_{i} +  3 k P_i  \left( \frac{\delta_{i,1}}{3} - P_1 \right) +\\ 
& k(\delta_{i,3} P_{i+1} + \delta_{i,2} P_{i-1}),
\end{aligned}
\end{equation}
for which $P_i=1/3$ is always a solution.  The corresponding matrix $J$ is identical to Equation~\ref{jaccase1} when $\chi=1/3$, and the linear stability analysis yields identical results to those in Equations~\ref{jaccase1}-\ref{lyap1}.  For example, if $\Gamma_i=1+a P_i^2$, we have a critical coupling of $a_c= a_{c,0}+ 6 k$, and when the coupling is fixed at $a=20$, synchronization is destroyed when $k>k_c=11/6$, which is consistent with numerical simulations (Figure~\ref{case2numericsA}).  In addition, at the transition point $\omega = \frac{\sqrt{3}}{6} (a_c+3)  = \sqrt{3} \left(2 + k_c \right)$ and the transition is always supercritical $\left( l_1=-\frac{3 \sqrt{3} (3+2 k)}{4+2k}<0\right)$.  Hence, while Equation~\ref{mfcase1} and Equation~\ref{mfcase2} correspond to microscopically distinct mechanisms and differ in higher order terms, when $\chi=1/3$ the linear stability properties of the fixed point $P_i=1/3$, which determine synchronization properties near the phase transition, are identical.  Intuitively, this correspondence arises because, in both cases, the division process redistributes the oscillators uniformly between the 3 possible states.

\begin{figure}
\includegraphics[width = 8.7 cm]{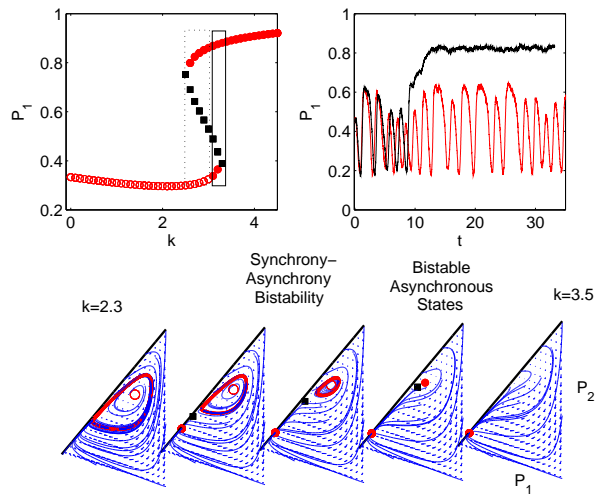}
\caption{Upper left panel: Steady state solution $P_1^*$ of the mean field model, Equation~\ref{mfcase2}. Stability is indicated by marker (open circles, unstable points; closed circles, stable points; squares, saddle points).  Dashed box is region where synchronous oscillations stably coexist with a non-oscillating state.  Solid box is region where two asynchronous states stably coexist.  Upper right panel: Two example time series of $P_1$ from numerical simulations with $N_0=2000$ starting from identical initial conditions, $P_1=0.45, P_2=0.45$. In one case, the system settles into stable oscillations (red).  In the other case, a fluctuation drives the system to the asynchronous fixed point following initial oscillations (black).  Bottom panel: phase portraits for $k=2.3, 2.6, 2.9, 3.2, 3.5$ (left to right).  Thin (blue) lines are example trajectories, thick (red) lines are stable limit cycles; stability of fixed points is indicated by markers (open (red) circles, unstable; closed (red) circles, stable; squares (black), saddle).  In all panels, $\chi=1$.}
\label{bistabfig2a}
\end{figure}

\begin{figure}
\includegraphics[width = 7.8 cm]{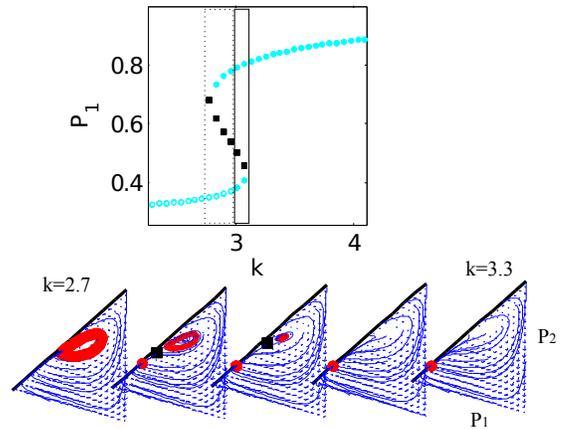}
\caption{Upper panel: Steady state solution $P_1^*$ of the mean field model, Equation~\ref{mfcase2}. Stability is indicated by marker (open circles, unstable points; closed circles, stable points; squares, saddle points).  Dashed box is region where synchronous oscillations stably coexist with a non-oscillating state.  Solid box is region where two asynchronous states stably coexist.  Bottom panel: phase portraits for $k=2.7, 2.85, 3.0, 3.15, 3.3$ (left to right).  Thin (blue) lines are example trajectories, thick (red) lines are stable limit cycles; stability of fixed points is indicated by markers (open (red) circles, unstable; closed (red) circles, stable; squares (black), saddle).  In all panels, $\chi=1$.}
\label{bistabfig2b}
\end{figure}

\begin{figure}
\includegraphics[width = 8.7 cm]{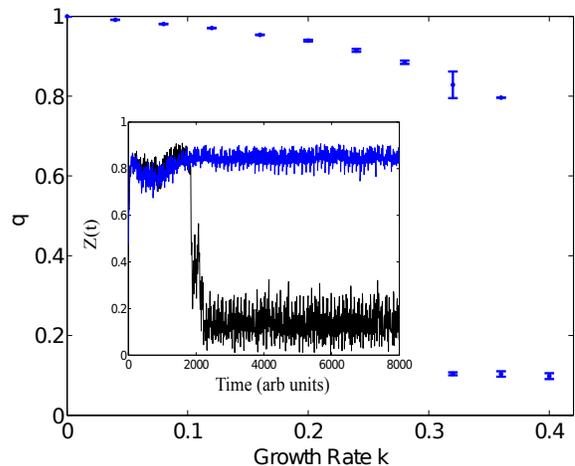}
\caption{Order parameter $q$ in growing population of identical Kuramoto oscillators, each with intrinsic frequency $\omega=10$.  Coupling strength $K= 1$; Number of initial oscillators $N_0=500$.  When population size reaches $N=10^6$, the population is reset to a size of $N=1000$ while preserving the distribution of oscillator phases. Points represent averages from simulations of 10 independent trials.  Insets:  Time series of order parameter $Z(t)$ for two typical simulations ($k=0.32$ in both cases) in the bistable region; the population stochastically switches between two stable states, one synchronized and one asynchronous.} 
\label{kuramotofig}
\end{figure}

For other values of $\chi$, however, state-dependent division can give rise to entirely new dynamics.  To explore this behavior, we performed numerical simulations for a wide range of parameters, as in Section~\ref{subseccase1}.  As a prototype model, we choose $\Gamma_i=1+a P_i^2$, but we later show that the similar behavior is observed for other coupling functions and even in continuous phase models.  While the amplitude of the oscillations decreases with increasing growth rate $k$ (Figure~\ref{case2numericsA}), the oscillations of $P_1$ and $P_2$ do not always occur around the symmetric values $(P_1, P_2) = (1/3, 1/3)$.  Furthermore, for sufficiently large values of $k$, the system appears to settle into a non-oscillating fixed point where the fraction of oscillators in states 1 and 2 can be significantly different.  As an example, for $\chi=2/3$ (Figure~\ref{case2numericsA}, upper right) the oscillations cease at $k \approx 3$, leading initially to a non-oscillating state where $P_1 <  P_2$.  This is somewhat counterintuitive, as only oscillators in state 1 divide, and the majority of daughter cells ($\chi=2/3$) are also in state 1.  As $k$ is further increased, the population is eventually dominated by oscillators in state 1, as one might expect.   Each individual oscillator continues to cycle through all three states, but on average, the distribution of states is not uniform.  

In addition, growth can dramatically affect the frequency of oscillations in the synchronous state, but the effect is no longer  monotonic for all values of $\chi$.  As $k$ is increased, we find oscillations of increasing frequency for $\chi=0$ and $\chi=1/3$, approximately constant frequency for $\chi=2/3$, and rapidly decreasing frequency for $\chi=1$ (Figure~\ref{case2numericsA}, lower panel inset).  It is somewhat surprising that the choice of $\chi$, alone, can lead to either an increase or decrease in the overall oscillation frequency.   Interestingly, the case $\chi=1$ also appears to undergo an abrupt transition for $k\approx 2.9$, indicating that the transition is qualitatively different from the supercritical Hopf bifurcation in the non-growing model. 

To examine this transition in detail, we performed linear stability analysis numerically for $\chi=1$.  In particular, for $\chi=1$, we find that the fixed point $P_1^*$ remains close to $P_1=1/3$ for $k<3$ and loses stability via a supercritical Hopf bifurcation at $k\approx 3$, in apparent contradiction with numerical results, which indicate a discontinuous transition at a smaller value of $k$ (see Figure~\ref{case2numericsA}). Interestingly, a more thorough numerical analysis reveals that a second branch of stable solutions arises at $k \approx 2.5$ (Figure~\ref{bistabfig2a}, upper left).  The emergence of this solution branch leads to two types of novel bistable behavior and underlies the abrupt transition observed in numerical simulations.  As $k$ is increased above $k \approx 2.5$, one finds bistability between a synchronized oscillating state and a non-oscillating state with $P_1>P_2>P_3$.  Surprisingly, the synchronized oscillations occur only when oscillator states are initially distributed within a relatively small section of phase space.  Initial conditions with $P_i=1$ for $i=$1, 2, or 3, for example, will (counterintuitively) lead to an asynchronous, non-oscillating steady state dominated by oscillators in state 1.  For finite populations, fluctuations may also lead the population to stochastically jump from one stable state (oscillations) to the other (fixed point) (Figure~\ref{bistabfig2a}, upper right).  As $k$ is further increased, the unstable fixed point at $P_1 \approx 1/3$ becomes stable (supercritical Hopf), leading to a small region of bistability between two non-oscillating states.  Finally, at $k \approx 3.4$, the lower branch of the solution disappears and the population settles into a fixed point on the upper branch of the solution curve.  Similar behavior occurs for other $\chi$ values in the approximate range $0.8 \le \chi \le 1$;  for smaller $\chi$, the Hopf bifurcation occurs prior to the emergence of the upper solution branch, leading to a larger region of bistability between asynchronous fixed points.

\subsubsection{Bistability in Other Classes of Oscillators}
When oscillators in any state can divide (i.e. Case 1), our linear stability analysis shows that similar behavior exists for a class of coupling functions, as long as the derivatives follow certain restrictions (see Section~\ref{subseccase1}).  Unfortunately, because the rotational symmetry of the model is broken when only one state can divide (i.e. Case 2), we are not able to provide similar analytical evidence that the bistable behavior discussed in Section~\ref{subseccase2} (e.g. Figure~\ref{bistabfig2a}) occurs for other choices of coupling function.  However, it is straightforward to perform numerical linear stability analysis for any particular model.

To confirm that these findings are not specific to the chosen form of the coupling function, we performed linear stability analysis for $\Gamma_i = e^{a P_i}$ at $a=3.65>a_{c,0}$.  As shown in Figure~\ref{bistabfig2b}, we see similar dynamics for this choice of coupling.  Specifically, for small $k$, we see stable oscillations.  As $k$ increases, we see a region of bistability between synchronous oscillations and an asynchronous fixed point, followed by a region of bistability between two asynchronous fixed points.  Finally, at $k \approx 3$, the lower branch of the solution disappears and the population settles into a fixed point on the other branch of the solution curve.  Numerically, we find that similar bistabilities also occur for other coupling functions, including multiple examples of the form $\Gamma_i = 1 + a P_i^n$ with $1<n\le 4$ (results not shown).  As before, these coupling functions lead to a supercritical Hopf bifurcation in the absence of growth.

Our analysis of discrete phase oscillators suggests that population growth can induce bistability between stable, asynchronous fixed points and stable, synchronous oscillations when the growth is strongly phase-dependent.  While the exact mathematical correspondence between these discrete phase models and classic continuous phase models, such as the Kuramoto model, has yet to be rigorously established, one might expect similar bistability to occur in continuous phase models as well.  To explore this question, we performed numerical simulations of populations of Kuramoto oscillators~\cite{kuramoto}, whose (continous) phases $\phi$ evolve according to
\begin{equation}
\label{kuramoto}
\dot{\phi}_i=\omega_i + \frac{K}{N} \sum_j \sin(\phi_j-\phi_i),
\end{equation}
$\omega_i$ is the intrinsic frequency of oscillator $i$, $K$ is a coupling parameter, and the sum runs over all oscillators in the population.  We take $\omega_i=\omega$ for all oscillators; for identical oscillators, a synchronous state exists for all $K>0$.  To incorporate phase dependent population growth, at each time step we allow oscillators whose phase falls in a given range, $\phi_0 \le \phi \le \phi_1$, to reproduce with probability per unit time $k$.  Following division, the daughter oscillator has a phase that is chosen to fall in the range $\phi_0 \le \phi \le \phi_1$ with uniform probability.  To circumvent numerical limitations due to exponentially growing populations, we start with $N_0=500$ oscillators and allow them to grow to a total size of $N=10^6$;  when N reaches this maximum value, we reset the population size to $N=1000$ while preserving the phase distribution of oscillators.  In practice, we preserve the approximate phase distribution by binning oscillator phases into a total of $M$ bins over the range $0 \le \phi \le 2 \pi$.  Prior to resetting the population size, we calculate the fraction of oscillators in each bin and choose the phases in the smaller population so that the distribution is conserved.  We find that similar behavior is observed as long as $M$ is sufficiently large.  In what follows, we choose $M=10$, $\phi_0 = \pi/3$, $\phi_1=2 \pi / 3$, and we set $K=1$.  

As with the discrete oscillators, numerical simulations suggest that population growth decreases the synchrony of the population, eventually leading to a discontinuous transition to asynchronous behavior and a region of bistability between synchronous and asynchronous oscillations (Figure~\ref{kuramotofig}).  Because the simulations involve finite populations, it is common to see trajectories that stochastically switch between stable oscillations and asynchronous behavior (Figure~\ref{kuramotofig}, inset).  While a full analysis of continuous models is an exciting avenue for future work, we stress that our goal here is simply to provide evidence that growth-induced bistability can occur in continuous phase models.  Further analysis along these lines is necessary and would be welcome, but it is beyond the scope of the current work.

\section{Excitable Elements}
\subsection{Model for Growing Populations of Excitable Elements}
\label{reviewexcite}
The results of Section~\ref{oscillators3state} raise the question of whether population growth might have similar effects on populations of coupled excitable elements. To explore this question in a simple context, we will model an excitable system as a discrete $m$-state system comprised of a quiescence state ($0$), an excited state ($1$), and a finite set of refractory states ($2, 3, ..., m-1$) using the model introduced in~\cite{rozenblit}. For each state $i$, the discretized phase is $\theta=2\pi i/m$. Here we study a simple four state system, which contains the minimum number of refractory states (2) required for stable synchronization~\cite{girvan}.

This discrete time model involves deterministic transitions between states $1\to2$ and $2\to3$ and probabilistic transitions from the quiescent state 0 to the activated state 1.  Coupling between elements is achieved by allowing the transition from state 0 to state 1 to depend on the states of the neighboring systems and a parameter $\sigma$ that measures the strength of the inter-element coupling. The last transition, from state 3 back to the quiescent state 0, is also probabilistic and occurs with probability $p_\gamma$, which is the same for all excitable elements~\cite{rozenblit}.  A thorough analysis of this model~\cite{rozenblit} reveals that, in the thermodynamic limit, the population transitions from an absorbing state to an active state at a critical value of $\sigma$.  More interestingly, for certain choices of $p_\gamma$, it can undergo synchronous oscillations within a range of coupling strengths $\sigma_1 < \sigma < \sigma_3$, and there exists a region of bistability between synchronous oscillations and an asynchronous fixed point for $\sigma_2 < \sigma < \sigma_3$, with $\sigma_1 < \sigma_2 < \sigma_3$.  

In what follows, we study this model in the case where individual elements are allowed to reproduce, producing daughter elements in potentially different states.  As with the oscillators, we focus on two cases:  all oscillators are equally likely to divide (case 1), and division occurs only for oscillators in one particular state (case 2). As before, the daughter element will be in the same state as the mother with probability $\chi$, and will be randomly assigned to each of the other states with probability $(1-\chi)/3$.  The discrete nature of the model makes it amenable to rapid numerical simulations~\cite{rozenblit}, which we explore in the next section.

\subsection{Numerical Simulations}
\label{numsimsE}
We performed discrete time simulations starting from $N_0=10^4$ globally coupled excitable elements for $t=500$ total time steps.  To avoid limits due to computational memory, we reduce the total population size by a factor of 5 (while preserving the state distribution) when the number of units reaches $5\times10^4$. As in~\cite{rozenblit}, we take the probability of exciting a quiescent state to be
\begin{equation}
p_\mu(t)=1-\left(1-\frac{\sigma}{N(t)} \right)^{N_{t}(1)}, 
\end{equation}
where $N(t)$ is the total number of units at time step $t$, $N_{t}(1)$ is the number of elements in state $1$ at time step $t$, each excited element can activate a neighbor in the quiescent state with probability $\sigma/N(t)$, and $p_\mu$ reflects the probability that the quiescent state is excited by at least one of its $N_{t}(1)$ active neighbors. We choose initial conditions so that $P_{0}(0)=0.8$ and $P_{0}(1)=0.2$, where $P_i(j)$ is the fraction of oscillators in state $j$ at time step $i$. For each choice of $k$ and $\chi$, we continuously increased coupling strength $\sigma$ from $2$ to $50$ for each successive simulation, using the steady state from the previous value of $\sigma$ as initial conditions for the next value. We then repeated the simulations starting from $\sigma=50$ and decreasing $\sigma$ to 0; when simulations reached a steady state, we calculated the order parameter $q$, which is the same as that used for oscillators (see Section~\ref{oscillators3state}).

\begin{figure}
\includegraphics[width = 7.cm]{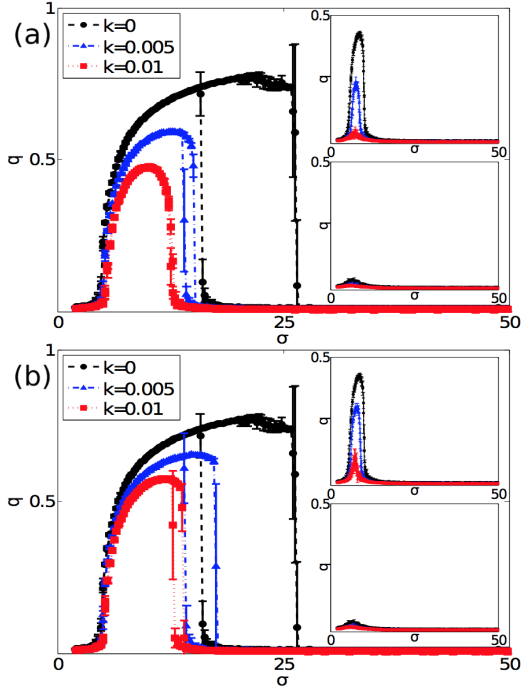}
\caption{Order parameter $q$ vs coupling strength $\sigma$ when all states divide for (a) $\chi=0$ and (b) $\chi=1$. Error bars are from standard deviations over $10$ runs. Black dashed, blue dash dot and red dotted line represent growth rate $k=0$, $0.005$ and $0.01$, respectively, at $p_\gamma=0.94$. For smaller values of $p_\gamma$, the size of the active region shrinks and the bistable area disappears, even in the absence of growth (upper insets in (a) and (b): $p_\gamma=0.84$). If $p_\gamma$ is further reduced, stable oscillations will eventually disappear (lower insets in (a) and (b): $p_\gamma=0.74$).}
\label{order1}
\end{figure}

In the case where all states can divide, the most salient effect of population growth is to decrease the range of $\sigma$ over which oscillations are stable (Figure~\ref{order1}).  For example, when $p_\gamma=0.94$, \cite{rozenblit} showed that the model undergoes a synchronizing transition as $\sigma$ is increased from zero.  Further increase of $\sigma$ leads to discontinuous re-entrant transition that includes a region of bistability between synchronous and asynchronous states (black curves, main panel).  When including population growth, we find that as $k$ is increased from zero, the size of both the synchronized and bistable regions shrink (Figure~\ref{order1}, blue curves); as $k$ is further increased, the bistable region eventually disappears (Figure~\ref{order1}, red curves) and, eventually, the entire synchronization region is lost (not shown). We observe a similar decrease in the size of the oscillatory regime for $\chi=0$ (top panels) and $\chi=1$ (bottom panels) and for smaller values of $p_\gamma$ where bistability does not exist, even in the absence of growth.  In all cases, increasing $\chi$ slightly counteracts the effect of population growth; that is, larger $\chi$ leads to a slightly larger region of synchrony and/or bistability.
        
\begin{figure}
\includegraphics[width = 7.cm]{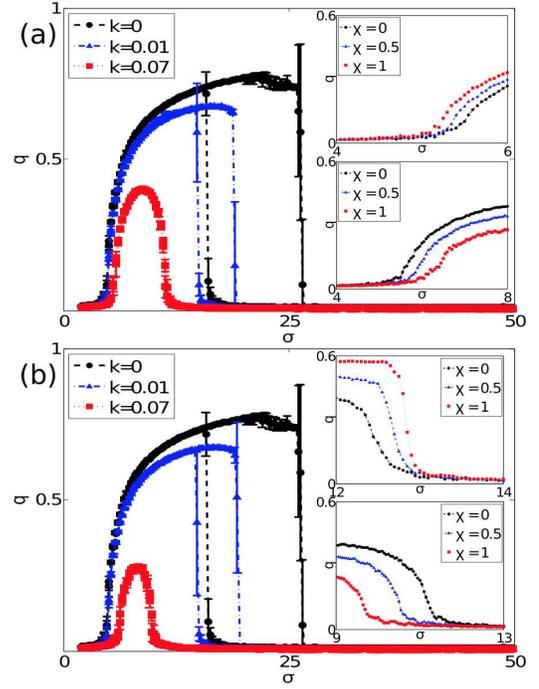}
\caption{Order parameter $q$ vs coupling strength $\sigma$ when division is restricted to one state.  Main panels: Division is restricted to units in state 1: (a) $\chi=0$ and (b) $\chi=1$. Black dashed, blue dash dot and red dotted lines represent growth rate $k=0$, $0.01$ and $0.07$, respectively at $p_\gamma=0.94$. When increasing $\chi$ from $0$ to $1$, the size of the active region shrinks (lower insets of both panels, $k=0.07$).  For comparison, upper insets show effect of increasing $\chi$ when division is independent of state (upper insets in (a) and (b), $k=0.01$).}
\label{order2}
\end{figure}

We find qualitatively similar behavior when division is restricted to only one state, such as state 1 (Figure~\ref{order2}).  Interestingly, however, we find that the effect of increasing $\chi$ will depend on which state is chosen for division.  Specifically, when division is restricted to states 0, 2 or 3, increasing $\chi$ will lead to an increase in the size of the active region (Figure~\ref{order2}, upper insets), as in the case where all states can divide (Figure~\ref{order1}).  By contrast, when division is restricted to state 1, increasing $\chi$ will lead to a decrease in the size of the active region (Figure~\ref{order2}, main panels and lower insets).

\begin{figure}
\includegraphics[width = 7.cm]{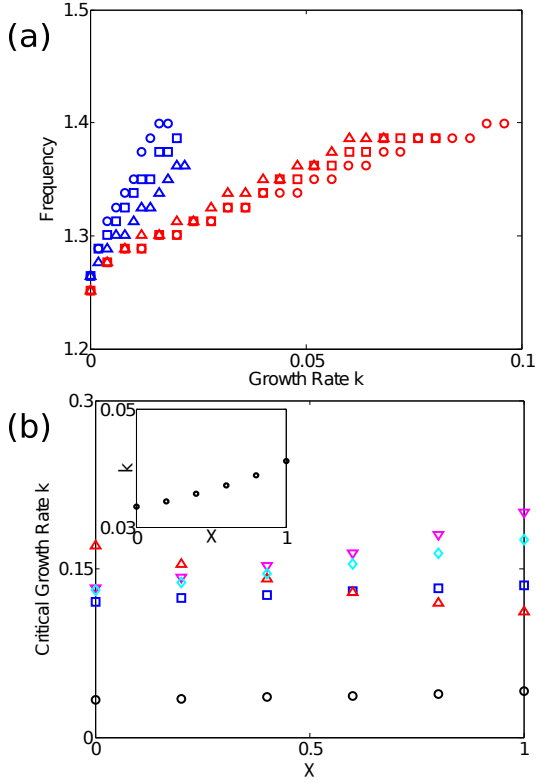}
\caption{Frequency and critical growth rate vs $\chi$. Upper figure: circle, square and up triangle represent $\chi=0$, $0.5$ and $1$, respectively. Blue color shows the case in which all states divide and red color means only state $1$ divides. $p_{\gamma}=0.9$ and $\sigma=8$. Lower figure: black circle, blue square, red up triangle, magenta down triangle and cyan diamond represent five cases (all states divide, only state 0 divides...), respectively. Inset: zoom in black circle (all states divide case)}
\label{freq}
\end{figure}

For all simulations, we also calculated 1) the frequency of macroscopic oscillations in the active regions and 2) the maximum growth rate--which we refer to as the critical growth rate--that still allows for synchronous oscillations. Figure~\ref{freq}(a) shows that population growth increases the frequency of oscillations, regardless of which state is chosen for division. Figure~\ref{freq}(b) illustrates that as $\chi$ increases, the critical growth rate increases when division is independent of state or is restricted to states 0, 2, or 3;  the opposite trend exists when division is restricted to state 1 (upper red triangles).

\subsection{Mean Field and Linear Stability Analysis}
\label{mfE}
To gain a systematic picture of these results, we follow the approach in~\cite{rozenblit} to develop a mean field approximation and derive full phase diagrams for these excitable systems.  Specifically, by assuming that probability of exciting a quiescent state via one of  its $N(t)-1$ neighbors is $\sigma P_{t}(1)/(N(t)-1)$, the probability of excitation via at least one excited neighbor is  
\begin{equation}
p_{\mu}(t)=1-\left(1-\frac{\sigma P_{t}(1)}{N(t)-1} \right)^{(N(t)-1)}.
\end{equation}
In the thermodynamic limit, $p_{\mu}$ becomes $1-e^{-\sigma P_{t}(1)}$. In the case where all states divide, the system is governed by the difference equations 
\begin{equation}
\begin{split}
&N_{t+1} (0)= p_{\gamma}N_t (3)+e^{-\sigma P_{t}(1)}N_{t}(0)\\
&\qquad\qquad+k(\chi N_{t}(0)+\frac{1-\chi}{3}(N_{t}(1)+N_{t}(2)+N_{t}(3)))\\
&N_{t+1} (1)= (1-e^{-\sigma P_{t}(1)})N_{t}(0)\\
&\qquad\qquad+k(\chi N_{t}(1)+\frac{1-\chi}{3}(N_{t}(0)+N_{t}(2)+N_{t}(3)))\\
&N_{t+1} (2)= N_{t}(1)\\
&\qquad\qquad+k(\chi N_{t}(2)+\frac{1-\chi}{3}(N_{t}(0)+N_{t}(1)+N_{t}(3)))\\
&N_{t+1} (3)=N_{t}(2)+(1-p_{\gamma})N_{t}(3)\\
&\qquad\qquad+k(\chi P_{t}(3)+\frac{1-\chi}{3}(N_{t}(0)+N_{t}(1)+N_{t}(2))),
\end{split}
\end{equation}
and the total number of oscillators follows $N(t+1)=(1+k)N(t)$.  If all equations are divided by the total number of oscillators, we arrive at equations for the probability $P_t(i)$ to be in state $i$ at time $t$, 
\begin{equation}
\begin{split}
\label{zero1}
&P_{t+1} (0)= \frac{p_{\gamma}}{1+k}P_t (3)+\frac{e^{-\sigma P_{t}(1)}}{1+k}P_{t}(0)\\
&\qquad\qquad+\frac{k}{1+k}(\chi P_{t}(0)+\frac{1-\chi}{3}(1-P_{t}(0)))
\end{split}
\end{equation}
\begin{equation}
\begin{split}
\label{one1}
&P_{t+1} (1)= \frac{1-e^{-\sigma P_{t}(1)}}{1+k}P_{t}(0)\\
&\qquad\qquad+\frac{k}{1+k}(\chi P_{t}(1)+\frac{1-\chi}{3}(1-P_{t}(1)))
\end{split}
\end{equation}
\begin{equation}
\begin{split}
&P_{t+1} (2)= \frac{1}{1+k}P_{t}(1)\\
&\qquad\qquad+\frac{k}{1+k}(\chi P_{t}(2)+\frac{1-\chi}{3}(1-P_{t}(2)))
\end{split}
\end{equation}
\begin{equation}
\begin{split}
\label{three1}
&P_{t+1} (3)=\frac{1}{1+k}P_{t}(2)+\frac{1-p_{\gamma}}{1+k}P_{t}(3)\\
&\qquad\qquad+\frac{k}{1+k}(\chi P_{t}(3)+\frac{1-\chi}{3}(1-P_{t}(3))).
\end{split}
\end{equation}
Similarly, if division is restricted to only one state, such as state 1, we have 
\begin{equation}
\begin{split}
\label{zero2}
&P_{t+1} (0)= \frac{p_{\gamma}}{1+kP_t(1)}P_t (3)+\frac{e^{-\sigma P_{t}(1)}}{1+kP_t(1)}P_t(0)\\
&\qquad\qquad+\frac{k(1-\chi)}{3(1+kP_t(1))}P_t(1)
\end{split}
\end{equation}
\begin{equation}
P_{t+1} (1)= \frac{1-e^{-\sigma P_{t}(1)}}{1+kP_t(1)}P_{t}(0)+\frac{k\chi}{1+kP_t(1)}P_t(1)
\end{equation}
\begin{equation}
P_{t+1} (2)= \frac{1}{1+kP_t(1)}P_{t}(1)+\frac{k(1-\chi)}{3(1+kP_t(1))}P_t(1)
\end{equation}
\begin{equation}
\begin{split}
&P_{t+1} (3)=\frac{1}{1+kP_t(1)}P_{t}(2)+\frac{1-p_{\gamma}}{1+kP_t(1)}P_{t}(3)\\
&\qquad\qquad+\frac{k(1-\chi)}{3(1+kP_t(1))}P_t(1).
\end{split}
\end{equation}
It is easy to derive similar equations when division is restricted to one of the other states. In all cases, nontrivial steady states $P^*(i)$ occur when $P_{t+1}(i)=P_{t}(i)$ ($i=0,1,2,3$). Since the probabilities are normalized to $1$, we can omit equations~(\ref{zero1}) and~(\ref{zero2}) and reduce each set of equations by one. 

To study the stability of nontrivial solutions, we linearize the equations near each solution; the corresponding Jacobian matrix $J$ when all states can divide is given by
\begin{equation}
J=\frac{J_N}{1+k}+\left(\frac{k}{1+k}\left(\chi-\frac{1-\chi}{3}\right)\right)I,
\end{equation}
where $J_N$ is the Jacobian for non-dividing populations~\cite{rozenblit}
\begin{equation}
J_N=\left. \left(
\begin{array}{ccc}
a_{11} & e^{-\sigma P_1}-1 & e^{-\sigma P_1}-1\\
1 & 0 & 0\\
0 & 1 & 1-p_\gamma\\
\end{array}
\right)\right|_{\vec{P}^*},
\end{equation}
$I$ is identity matrix, and $a_{11}=\sigma e^{-\sigma P_1}(1-P_1-P_2-P_3)+e^{-\sigma P_1}-1$. At steady state, $P_1^*=P_2^*=p_\gamma P_3^*$. We note that even when $\chi=1$ and the excitable elements reproduce to form identical daughter cells, the model does not reduce to the corresponding non-growing model.  Hence, unlike the oscillator model, growth will modify the dynamics of these excitable systems even in the case when all states can divide and $\chi=1$.  For cases where division is restricted to one state, the Jacobian can be readily calculated numerically, but it cannot be simply written in terms of $J_N$.  

\begin{figure}
\includegraphics[width = 6.cm]{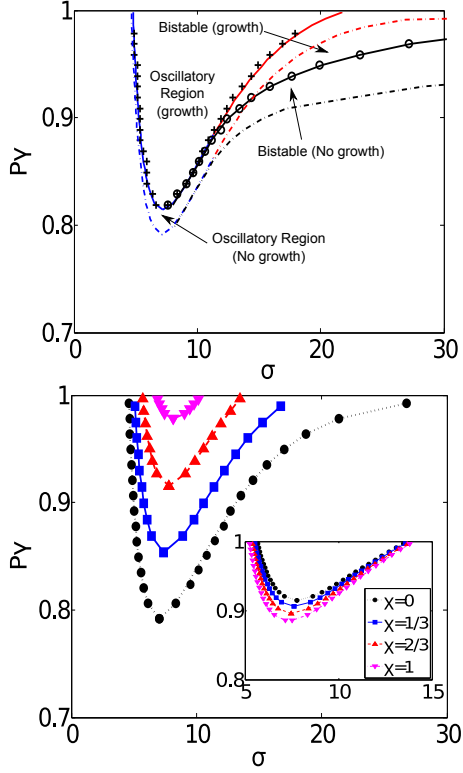}
\caption{Phase diagrams when all oscillator states can divide.  Upper panel:  Phase diagram $p_{\gamma}$ vs $\sigma$ when all states divide and $k=0.005$, $\chi=1$.   Curves indicate nature of bifurcation (blue, supercritical Hopf; red, subcritical Hopf; black, global saddle node of limit cycles).  Regions of bistability between synchronous and asynchronous states are indicated.  Crosses and circles indicate results from numerical simulations ($N_0=10^4$). Black line shows numerical $\sigma_{c3}$ from mean field analysis. Cross and circle represent simulation results. Dash dot means non-growth for comparison.  Lower panels: phase boundaries (excluding bistable regions) when all states divide, $\chi=0$; $k=0$ (black circle), $k=0.01$ (blue square), $k=0.02$ (red up triangle), $k=0.03$ (magenta down triangle). Lower right inset, $k=0.02$ for different values of $\chi$:  Black circle, blue square, red up triangle, magenta down triangle represent $\chi=0$, $\frac{1}{3}$, $\frac{2}{3}$ and $1$, respectively. }
\label{phase}
\end{figure}

\begin{figure}
\includegraphics[width = 8.4 cm]{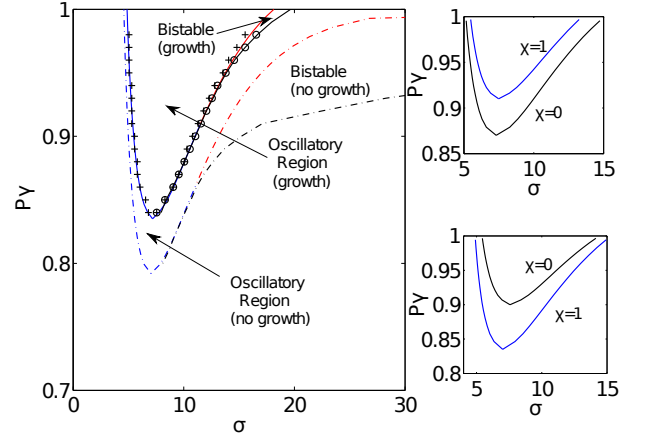}
\caption{Phase diagrams when division is restricted to one state.  Main panel: phase diagram for excitable elements with growth (solid lines; $\chi=0$, $k=0.04$, only state 1 can divide) and without growth (dashed lines).  Curves indicate nature of bifurcation (blue, supercritical Hopf; red, subcritical Hopf; black, global saddle node of limit cycles).  Regions of bistability between synchronous and asynchronous states are indicated.  Crosses and circles indicate results from numerical simulations ($N_0=10^4$). Right panel, phase boundaries (excluding bistable regions) for $\chi=0$ and $\chi=1$ when only state 1 (top) or only state 2 (bottom) divides; $k=0.07$ in both cases.}
\label{comparison}
\end{figure}
 
As in the case of continuous time systems, the eigenvalues provide information about the stability of each fixed point.  Specifically, bifurcations between stable and unstable fixed points occur when $|\lambda|=1$, allowing us to dilineate phase boundaries separating oscillatory and non-oscillatory regimes.  Specifically, we are interested in the location of Neimark-Sacker (NS) bifurcations, which are the discrete time analog of a Hopf bifurcation. As with Hopf bifurcations, the nature of the transition is given by the sign of the first Lyapunov coefficient:  the NS bifurcation is supercritical if $l_1<0$ and subcritical if  $l_1>0$.  Following standard bifurcation theory (see, for example,~\cite{rozenblit}), we calculate $l_1$ as
\begin{equation}
\begin{split}
&l_1=\frac{1}{2}Re\left\{\bar{\lambda}\left[\left\langle\vec{s},\vec{C}(\vec{r}, \vec{r}, \bar{\vec{r}})\right\rangle  \right.\right. \\
&\qquad+2\left\langle\vec{s},\vec{B}(\vec{r}, (I-J)^{-1}\vec B(\vec r,\bar{\vec r}))\right\rangle\\
&\qquad\left.\left. +\left\langle\vec{s},\vec{B}(\bar{\vec{r}},(\lambda^2I-J)^{-1}\vec B(\vec r,\vec r))\right\rangle \right]\right\},
\end{split}
\end{equation}
where brackets represent the standard complex inner product, $\vec r$ is the eigenvector of the Jacobian matrix with corresponding eigenvalue $\lambda$, $\vec s$ is the eigenvector of $J^T$ with eigenvalue $\lambda^*$, and normalization is chosen so that $\left\langle\vec{r},\vec{r}\right\rangle=1$ and $\left\langle\vec{s},\vec{r}\right\rangle=\sum_{i=1}^{3}\bar{s}_ir_i=1$. Furthermore, $\vec B(\vec x, \vec y)$ and $\vec C(\vec x, \vec y, \vec z)$ are vector-valued multi-linear functions 
\begin{equation}
\begin{aligned}
B_i(\vec x, \vec y)&=\sum_{j,k=1}^{3}\left.\frac{\partial^2 F_i(\vec P_t)}{\partial P_t(j) \partial P_t(k)}\right|_{\vec P^*, \sigma_c}x_j y_k, \\
C_i(\vec x, \vec y, \vec z)&=\sum_{j,k,l=1}^{3}\left.\frac{\partial^3 F_i(\vec P_t)}{\partial P_t(j) \partial P_t(k) \partial P_t(l)}\right|_{\vec P^*, \sigma_c}x_j y_k z_l,
\end{aligned}
\end{equation} 
where $\vec x=(x_1, x_2, x_3)^T$, $\vec y=(y_1, y_2, y_3)^T$ and $\vec z=(z_1, z_2, z_3)^T$ are arbitrary vectors.  As in~\cite{rozenblit}, when $l_1>0$ (indicating a subcritical bifurcation and the corresponding bistability), we supplement the above calculations with simulations of the mean field equations to determine the location of the bistable regimes, which are not fully determined by linear stability properties. 

Figure~\ref{phase} shows the resulting phase diagrams for the case where all states divide, and Figure~\ref{comparison} shows the phase diagram when division is restricted to state 1.  Population growth has several obvious effects on the excitable systems.  First, growth can shift the location of the fixed points, but unlike the oscillator case, we do not find evidence of new fixed points. Consistent with our simulations, growth shifts the phase boundaries to higher values of $p_{\gamma}$, leading to a smaller region of synchronized oscillations and a significantly reduced region of bistability; as $k$ is further increased, the oscillatory region is eventually eliminated (see Figure~\ref{phase}, bottom panel). The phase diagrams indicate that adding growth can have significant effects on the dynamics, depending on the values of $\sigma$ and $p_{\gamma}$.  Consider, for example, the main panel of Figure~\ref{comparison}.  Systems originally undergoing oscillations can enter a bistable state (e.g. $(\sigma,p_\gamma)=18, 0.98)$) or a non-oscillating active state (e.g. $(\sigma,p_\gamma)=8, 0.82)$) when growth is increased to $k=0.04$.  Qualitatively similar behavior is observed when division is restricted to one of the other states, or when all states can divide (Figure~\ref{phase}). 

Interestingly, we also find that the effect of $\chi$ on the phase diagram will depend on which state is chosen for division (Figure~\ref{phase}, bottom inset;  Figure~\ref{comparison}, right panels).  For example, increasing $\chi$ at a fixed value of $k$ will raise the phase boundary to higher $p_{\gamma}$ when state 1 divides, but will lower the boundary when states 0, 2, or 3 divide (Figure~\ref{comparison}, right panels) or when all states can divide (Figure~\ref{phase}, bottom inset). When only the excited state (1) divides, self-similarity between mother and daughter oscillators decreases the area of phase space over which synchrony can occur, while such self-similarity increases the size of the synchronized region when any of the non-excited states divide.

\section{Discussion}
\label{discussion}
We have shown that population growth, and the corresponding redistribution of oscillator phases, can induce a wide range of new dynamic behaviors in systems of coupled oscillators and excitable elements.  In particular, when growth is independent of oscillator phase, increasing growth rate leads to an increase in oscillation frequency and a decrease in phase synchrony, eventually culminating in a transition to a non-oscillating steady state.  Interestingly, the growth-induced transition can be subcritical, even when the non-growing model exhibits a supercritical bifurcation; in that case, one sees a bistable region with coexisting synchronous and asynchronous states, depending on the derivatives of the coupling function.  When division is strongly state dependent, growth can again lead to extinction of oscillations, but in certain parameter regimes, one finds new asynchronous states with unequal phase distributions, bistability between two asynchronous states or between asynchronous and oscillatory states, or modulation of the bulk oscillation frequency.  In excitable systems, which may include bistable behavior even in the absence of growth, the most salient effects of growth are to shift the phase boundaries separating active, oscillatory, and bistable regimes while also increasing the frequency of super-threshold oscillations.  In practice, the shifting phase boundaries can lead to a range of different dynamical effects, many of which mirror the behaviors seen in oscillators.  For example, systems originally undergoing oscillations can enter a bistable state or a non-oscillating active state when growth is increased.  

Several recent studies have focused on related ideas.  First, a number of studies have examined cell-density dependent synchronization (see, for example, ~\cite{elowitzPNAS, koseka}). In these systems, synchronization of intracellular dynamics can be modified at high cell densities as a result of biochemical communication known as quorum sensing.  These studies do not directly explore the effects of population growth rate, but instead treat cell density as the relevant control parameter that governs not only intracellular coupling, but also the dynamics of individual cells.  In addition, the authors of~\cite{david2} develop a detailed biochemical model of circadian clocks in growing cyanobacteria populations.  They show that otherwise stable oscillations--specifically, those driven by phosphorylation cycles--are destabilized at high growth rates via a supercritical Hopf bifurcation for the chosen range of parameters. Therefore, in fast growing populations, additional stabilizing mechanisms (transcription-translation cycles) are required to preserve integrity of the oscillations.  It would be interesting to see if subcritical bifurcations, including the bistability observed in our models, occur in different parameter regimes.  In bacteria populations, an elegant series of studies has recently laid the groundwork for synthetic sensors and logical programming in living systems based on tunable spatiotemporal oscillations in growing populations~\cite{hasty2010, hasty2014}.  However, the effects of growth rate on these dynamics are not addressed in detail.   Finally, recent theoretical work~\cite{emelianova} has demonstrated a rich collection of dynamical behaviors in small chains of coupled oscillators gradually increasing in number. By contrast, here we focus on large systems of oscillators and the corresponding phase transitions to macroscopic oscillations.  To our knowledge, this work is the first to systematically address how coupling between population growth and oscillator phase can effect synchronization.

We have shown that population growth can dramatically influence synchronization phenomena and, in some cases, lead to entirely new dynamical states in populations of coupled oscillators and excitable elements. While we focused on discrete phase models because of their relative simplicity, we hope these results will motivate future explorations on the interplay between synchronization and population growth in additional models of oscillators, excitable elements, and perhaps more general dynamical systems. Given the theoretical importance of self-synchronization in statistical physics and its ubiquity in biological systems, we believe that the potential effects of population growth on collective oscillations will prove to be an important and rich topic for future exploration.  

\bibliography{oscillatorsbib2014.bib}

\end{document}